# SHUFFLING:IMPROVING DATA SECURITY IN AD HOC NETWORKS BASED ON UNIPATH ROUTING

K. KARNAVEL[†]

Assistant Professor
Department of Computer Science and Engineering
Anand Institute of Higher Technology, Chennai
E-mail: treseofkarnavel@gmail.com

A. BALADHANDAYUTHAM

PG Student
Department of Computer Science and Engineering
Anand Institute of Higher Technology, Chennai
E-mail: sivabaladhandayutham@gmail.com

Abstract:
An ad hoc network is a self-organizing network with help of Access Point (AP) of wireless links connecting nodes to another. The nodes can communicate without infrastructure network. They form an random topology (BSS/ESS), where the nodes play the role of routers and are free to move arbitrarily. Ad hoc networks confirmed their efficiency being used in different fields but they are highly vulnerable to security attacks and dealing with this is one of the main challenges of these networks today. In recent times, a few solutions are proposed to provide authentication, confidentiality, availability, secure routing and intrusion prevention in ad hoc networks. Employ security in such dynamically changing networks is a hard task. Ad hoc network uniqueness must be taken into contemplation to be clever to design efficient clarification. Here we spotlight on improving the flow transmission confidentiality in ad hoc networks based on unipath routing. Definitely, we take advantage of the being of multiple paths between nodes in an ad hoc network to increase the confidentiality robustness of transmitted data with the help of Access Point. In our approach the original message to secure is split into shares through access point that are encrypted and combined then transmitted along different disjointed existing paths between sender and receiver. Still if an attacker succeeds to attain one or more transmitted contribute to the probability that the unique message will be reconstituted is very low.

## 1. Introduction

A wireless is a decentralized type of wireless network.[1] The network is ad hoc because it does not rely on a preexisting infrastructure, such as routers in wired networks or access points in managed (infrastructure) wireless networks. Instead, each node participates in routing by forwarding data for other nodes, so the determination of which nodes forward data is made dynamically on the basis of network connectivity. In addition to the classic routing, s can use flooding for forwarding the data. An typically refers to any set of networks where all devices have equal status on a network and are free to associate with any other device in link range. Often refers to a mode of operation of IEEE 802.11 wireless networks. Using cooperative wireless communications improves immunity to interference by having the destination node combine self-interference and other-node interference to improve decoding of the desired signal.

## 2. Literature review

- Jalel Ben Othman(2009)-implemented enhancing data security in ad-hoc networks based on multipath routing. In this, the original message to secure is split into shares that are encrypted and combined then transmitted along different disjoint existing paths between sender and receiver. Even if an attacker succeeds to obtain one or more transmitted shares, the probability that the original message will be reconstituted is very low[1].
- Liang Zhou( January-2010)-scheduled security critical multimedia applications in heterogeneous networks. it takes into account application's timing and security requirements in addition to precedence constraints. As it finds resource allocations heuristically it maximizes the quality of security and the probability of meeting deadlines for all the multimedia applications running on heterogeneous networks. if the distortion model constructed is not accurate scalable graph based application becomes failure thereby decreasing the flexibility and efficiency[2].
- William R Claycomb(March-2010)-implemented a node level security policy in wireless networks. It reduces significant attacks at the node levels through distribution model algorithm and identity based cryptography. As sensor networks continues to be part of everyday life security of the network is critical to maintain[3].
- Frank A. Zdarsky(April-2010)- improved security in wireless mesh backhaul(WMB) architecture. It resolves the underlying security issues for the cases that standard security solutions do not exist. As everything is based on basic assumptions WMB architecture is still susceptible to security vulnerabilities[4].
- Pavlo Bykovyy(November-2011)-proposed a specialized interface for detectors network of alarm system. It provides a high level of security inside the system and the cost is low as it reduces the cabling required. If the traditional detectors used in this system becomes failure then the entire system cannot be protected[5].
- Hannes Holm(December-2011)-improved the performance of automated network vulnerabilities using remediated security issues. It is independent of the system credentials used by the system. However manual effort is needed to provide complete accuracy and it is prone to false alarms in the networks[6].
- Aura Reggiani(2012)-improved network resilience for transport security by considering methodological reflections. through this resilience the system can absorb shocks without catastrophic changes in the functional organization and an effective tool in understanding the evolutionary paths of complex spatial networks. Yet the measure of resilience is still a critical issue as they still remain at a formal theoretical level [7].

## 3. Security in ad hoc networks

In mobile s, security depends on several parameters (authentication, confidentiality, integrity, non-repudiation and availability) .Without one of these parameters, security will not be complete. Without authentication, an attacker could masquerade a node, thus being able to have unauthorized access to the resources and to sensitive information. Confidentiality ensures that exchanged information will not be consulted by unauthorized nodes. Integrity means that information can only be modified by nodes allowed to do it and by their own willing. Non-repudiation permits obtaining a proof that information are sent or received by someone. Thus, a sender or a receiver cannot deny that he sent or received the concerned information. And finally, availability ensures that network services can survive despite any attack. s are exposed to many possible attacks. We can classify these attacks into two kinds: passive and active attacks. In passive attacks, attackers do not disrupt the operation of routing protocol but only attempt to discover valuable information by listening to the routing traffic. Defending against such attacks is difficult, because it is usually impossible to detect eavesdropping in a wireless environment. Furthermore, routing information can reveal relationships between nodes or disclose their IP addresses. If a route to a particular node is requested more often than to other nodes, the attacker will be able to expect that the node is important for the network, and disabling it could bring the entire network down. Unlike passive attacks, active attacks are often detectable. An active attack can mainly be: Blackhole attack, where a malicious node uses the routing protocol to advertise itself as having the shortest path to the node sending packets it wants to intercept. If the malicious reply reaches the requesting node before the reply from a correct node, a forged route is created. Thus the malicious node can do anything with the received packets. Another of active attacks is Routing tables overflow attacks. In this attacks, the attacker attempts to create routes to non-existent nodes. The goal is to create enough routes to prevent new routes from being created or to cause routing tables overflow. Sleep deprivation attacks is also an active attack where an attacker attempts to consume batteries by requesting routes, or by forwarding unnecessary packets. There are also location disclosure attacks which can reveal information about the node locations or the network structure and denial of service attacks (DoS), where an attacker can make the crashed or congested by different possible methods. Finally, we can cite impersonation attacks that we can avoid if node authentication is supported. Compromised nodes may be able to join the network while being undetectable or send false routing information masqueraded as trusted node. There were the most important and dangerous possible attacks in s. To protect s against different kinds of attacks and to solve security problems, many approaches were proposed.

### 3.1 Key management: Resurrecting duckling security policy

The basic concept in this approach is the use of master/slave relations between devices. Master and slave share a common secret. This association can be only broken by the master. Duckling (slave) will recognize as mother (master) the first entity sending him a secret key on a protected channel. This procedure is called Imprinting. It will obey always its mother, who says to him who is trusted nodes to whom it can speak (by subjecting the slave an access control checklist). If the link is stopped by the master with one of his slaves or if a network anomaly happened, the slave state becomes dead. It can be resurrected by accepting a new Imprinting operation. There is a hierarchy of master/slaves because a slave has the right to become master. The root is a person who controls all the devices. This solution is only effective for devices with weak processors and limited capacity.

### 3.2 Key agreement based password

The work developed in [7] draws up the scenario of a group wishing to provide a secured session in a conference room without supporting any infrastructure. The approach describes that these is a Weak Password that the entire group will share (for example by writing it on a board). Then, each member contributes to create a part of the session key using the weak password. This secured session key makes it possible to establish a secured channel without any centralized trust or infrastructure. This solution is adapted, to the case of conferences and meetings, where the number of nodes is small. It is rather strong solution since it does not have a strong shared key. But this model is not sufficient for more complicated environments. If we consider a group of people who do not know each other and want to communicate confidentially, this model becomes invalid. Another problem emerges if nodes are located in various places because then distribution of the Weak Password will not be possible any more.

### 3.3 Intrusion detection

One of recent interesting aspects of security in wireless networks, especially s is intrusion detection. It concerns detecting inappropriate, incorrect, or anomalous activity in the network. In [10] authors examine the vulnerabilities of wireless networks and argue that intrusion detection is a very important element in the security architecture for mobile computing environment. Intrusion prevention measures, such as encryption and authentication, can be used in s to reduce intrusion, but cannot eliminate them. In their architecture [10], they suggest that intrusion detection and response systems should be both distributed and cooperative to suit the needs of mobile s. Also, every node participates in intrusion detection and response. Thus there are individual IDS (Intrusion Detection Systems) agents placed on each node. They conclude that intrusion detection can compliment intrusion prevention techniques (such encryption, authentication, secure MAC, secure routing, etc.) to secure mobile computing environment.

### 3.4 Shuffling

The security for reliable data delivery scheme addresses data confidentiality and data availability in a hostile ad hoc environment. The confidentiality and availability of messages exchanged between the source and destination nodes are statistically enhanced by the use of unipath routing. At the source, messages are split into multiple pieces that are sent out via single independent path. The destination node then combines the received pieces to reconstruct the original message. The Shuffle scheme used shuffling the frame in order of sequence before transmitting the data. In the continuation to avoid intruder problem in ad hoc network Fig.1.

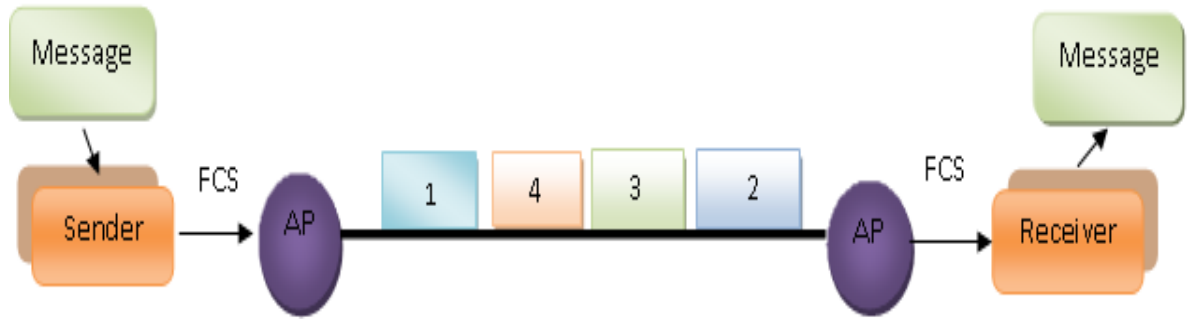

Fig: 1 ad hoc unipath routing process

### 4. Securing data based unipath routing(SDUP)

The idea behind our protocol is to divide the initial message into parts then to encrypt and combine these parts by pairs. Then we exploit the characteristic of existence of single paths between nodes in an to increase the robustness of confidentiality. This is achieved by sending encrypted combinations on the same path between the sender and the receiver. In our solution, even if an attacker succeeds to have one part or more of transmitted parts, the probability that the original message can be reconstructed is low.

*4.1 Paths selection in SDUP*

In the topology changes frequently, which makes wireless links instable. Sometimes packets might be dropped due to the bad wireless channel conditions, the collision at MAC layer transmission, or because of out of date routing information. When packet loss does occur, non-redundant share allocation will disable the reconstruction of the message at the intended destination. To deal with this problem, it is necessary to introduce some redundancy (if there is enough paths) in SDUP protocol to improve the reliability, (i.e. the destination would have better chance to receive enough shares for reconstructing the initial message). We propose that the decision of using or not redundancy will be taken according to the average mobility of the network's nodes and to the existing path number. SDUP is based on unipath routing in s. Routing in a presents great challenge because the nodes are capable of moving and the network topology can change continuously and unpredictably. Whereas we need that security will be the essential parameter in choosing different paths in such a way that the message security is maximized Fig.2.

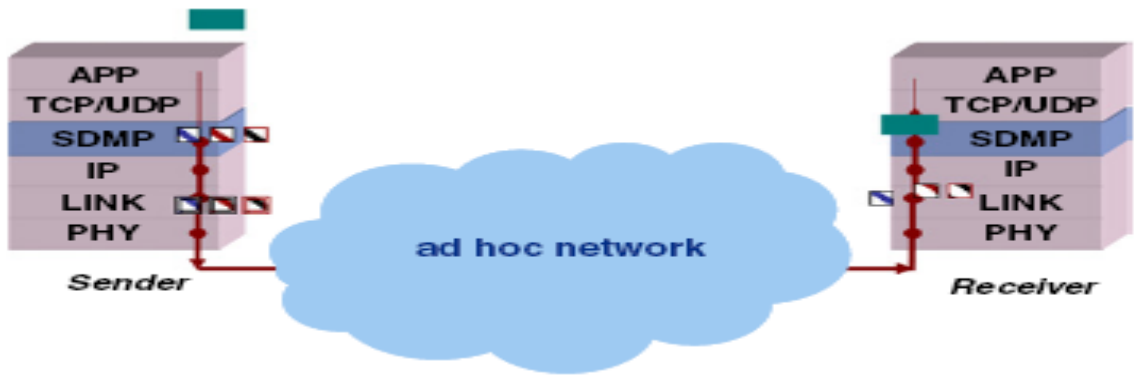

Fig: 2 Data Flow Architectur

### 5. Process Flowchart for collision Avoidance

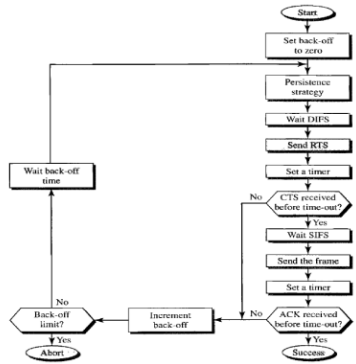

Fig:3 Process Flowchart

The flowchart exploring the flow of process in wireless link ,before exchange the frams and the distnace between stations can be great. Signal fading prevent a station at one end from hearing a collision at the other end Fig.3.

### 6. Conclusion

In this paper, we proposed a new solution that treats data confidentiality problem by exploiting a very important ad hoc network feature, which is used the existence of unipaths between nodes. Our proposal improves data security efficiently without being time-consuming. It is not complicated and can be implemented in different infrastructure network. This protocol is strongly based on unicast routing characteristics of ad hoc networks and uses a route selection based on security costs and BSS, ESS provide the high data security for secure data transmission without using access point (AP). In this continuation our process before sending the data to check the path is idle or busy, afterword's, to send the data. SDUP protocol can be combined with other solutions which consider other security aspects than confidentiality to improve significantly the efficiency of security systems in infrastructure network.